\documentclass[aps,prc,twocolumn,superscriptaddress,preprintnumbers,amsmath,amssymb,floatfix,showpacs,showkeys,nofootinbib]{revtex4}

\usepackage{graphicx, fink}
\usepackage{epsfig}
\usepackage{dcolumn}
\usepackage{bm}
\usepackage{longtable}
\usepackage{color}
\usepackage{float}
\usepackage{comment}
\usepackage{amsmath}
\usepackage{amsfonts}
\usepackage{xspace}
\usepackage{multirow}
\usepackage{nicefrac}
\usepackage[normalem]{ulem}
\usepackage{siunitx}

\usepackage{tikz}
\usetikzlibrary{decorations.pathmorphing}
\usetikzlibrary{arrows}
\usetikzlibrary{decorations.markings}
\usetikzlibrary{calc}
\usetikzlibrary{patterns}

\usepackage{textcomp}  

\usepackage{array}
\newcolumntype{L}[1]{>{\raggedright\let\newline\\\arraybackslash\hspace{0pt}}m{#1}}
\newcolumntype{C}[1]{>{\centering\let\newline\\\arraybackslash\hspace{0pt}}m{#1}}
\newcolumntype{R}[1]{>{\raggedleft\let\newline\\\arraybackslash\hspace{0pt}}m{#1}}

\begin{document}

\newcommand{\ChiEFT}{\ensuremath{\chi}EFT\xspace}

\preprint{MAX-lab $^{2}$H$(\gamma,\gamma)$ Summary Article for arXiv}

\title{Compton Scattering from the Deuteron above Pion-Production Threshold}

\author{\mbox{B.~Strandberg}}
\altaffiliation{Present address: Nikhef, Science Park 105, 1098 XG Amsterdam, Netherlands}
\affiliation{School of Physics and Astronomy, University of Glasgow, Glasgow G12 8QQ, Scotland UK}

\author{\mbox{J.~R.~M.~Annand}}
\affiliation{School of Physics and Astronomy, University of Glasgow, Glasgow G12 8QQ, Scotland UK}

\author{\mbox{W.~Briscoe}}
\affiliation{Institute for Nuclear Studies, Department of Physics, The George Washington University, Washington DC 20052, USA}

\author{\mbox{J.~Brudvik}}
\affiliation{MAX IV Laboratory, Lund University, SE-221 00 Lund, Sweden}

\author{\mbox{F.~Cividini}}
\affiliation{Institut f{\"ur} Kernphysik, Johannes Gutenberg-Universit{\"at} Mainz, D-55099 Mainz, Germany}

\author{\mbox{L.~Clark}}
\affiliation{School of Physics and Astronomy, University of Glasgow, Glasgow G12 8QQ, Scotland UK}

\author{\mbox{E.~J.~Downie}}
\affiliation{Institute for Nuclear Studies, Department of Physics, The George Washington University, Washington DC 20052, USA}

\author{\mbox{K.~England}}
\affiliation{College of Engineering, University of Massachusetts Dartmouth, North Dartmouth MA 02747, USA}

\author{\mbox{G.~Feldman}}
\affiliation{Institute for Nuclear Studies, Department of Physics, The George Washington University, Washington DC 20052, USA}

\author{\mbox{K.~G.~Fissum}}
  \altaffiliation{Corresponding author; \texttt{kevin.fissum@nuclear.lu.se}}
\affiliation{Department of Physics, Lund University, SE-221 00 Lund, Sweden}

\author{\mbox{D.~I.~Glazier}}
\affiliation{School of Physics and Astronomy, University of Glasgow, Glasgow G12 8QQ, Scotland UK}

\author{\mbox{K.~Hamilton}}
\affiliation{School of Physics and Astronomy, University of Glasgow, Glasgow G12 8QQ, Scotland UK}

\author{\mbox{K.~Hansen}}
\affiliation{MAX IV Laboratory, Lund University, SE-221 00 Lund, Sweden}

\author{\mbox{L.~Isaksson}}
\affiliation{MAX IV Laboratory, Lund University, SE-221 00 Lund, Sweden}

\author{\mbox{R.~Al~Jebali}}
\affiliation{School of Physics and Astronomy, University of Glasgow, Glasgow G12 8QQ, Scotland UK}

\author{\mbox{M.~A.~Kovash}}
\affiliation{Department of Physics and Astronomy, University of Kentucky, Lexington, KY 40506, USA}

\author{\mbox{S.~Lipschutz}}
\altaffiliation{Present address: National Superconducting Cyclotron Laboratory, Michigan State University, Michigan 48824, USA}
\affiliation{School of Physics and Astronomy, University of Glasgow, Glasgow G12 8QQ, Scotland UK}

\author{\mbox{M.~Lundin}}
\affiliation{MAX IV Laboratory, Lund University, SE-221 00 Lund, Sweden}

\author{\mbox{M.~Meshkian}}
\affiliation{Department of Physics, Lund University, SE-221 00 Lund, Sweden}

\author{\mbox{D.~G.~Middleton}}
\affiliation{Institut f{\"ur} Kernphysik, Johannes Gutenberg-Universit{\"at} Mainz, D-55099 Mainz, Germany}
\affiliation{Mount Allison University, Sackville, New Brunswick E4L 1E6, Canada.}

\author{\mbox{L.~S.~Myers}}
\affiliation{Bluffton University, 1 University Drive, Bluffton, OH 45817, United States}

\author{\mbox{D.~O'Donnell}}
\affiliation{School of Physics and Astronomy, University of Glasgow, Glasgow G12 8QQ, Scotland UK}

\author{\mbox{G.~O'Rielly}}
\affiliation{College of Engineering, University of Massachusetts Dartmouth, North Dartmouth MA 02747, USA}

\author{\mbox{B.~Oussena}}
\affiliation{Institute for Nuclear Studies, Department of Physics, The George Washington University, Washington DC 20052, USA}

\author{\mbox{M.~F.~Preston}}
\affiliation{Department of Physics, Lund University, SE-221 00 Lund, Sweden}

\author{\mbox{B.~Schr\"oder}}
  \affiliation{MAX IV Laboratory, Lund University, SE-221 00 Lund, Sweden}
\affiliation{Department of Physics, Lund University, SE-221 00 Lund, Sweden}

\author{\mbox{B.~Seitz}}
\affiliation{School of Physics and Astronomy, University of Glasgow, Glasgow G12 8QQ, Scotland UK}

\author{\mbox{I.~Strakovsky}}
\affiliation{Institute for Nuclear Studies, Department of Physics, The George Washington University, Washington DC 20052, USA}

\author{\mbox{M.~Taragin}}
\affiliation{Institute for Nuclear Studies, Department of Physics, The George Washington University, Washington DC 20052, USA}

\collaboration{The COMPTON@MAX-lab Collaboration}
\noaffiliation
 \date{\today}

\begin{abstract}

  The electromagnetic polarizabilities of the nucleon are fundamental nucleon-structure observables that characterize its response to external electromagnetic fields.
The neutron polarizabilities can be accessed from Compton-scattering data on light nuclear targets. Recent measurements of the differential cross section for Compton scattering on the deuteron below the pion-production threshold have decreased the uncertainties in the neutron polarizabilities, yet the proton polarizabilities remain known substantially more accurately. As the sensitivity of the cross section to the polarizabilities increases with incident photon energy, measurements above the pion threshold may offer a way for an improved determination of the neutron polarizabilities. In this article, the first measurement of the cross section for Compton scattering on the deuteron above the pion-production threshold is presented.


\end{abstract}


\maketitle

Nucleon polarizabilities are fundamental nucleon-structure observables that characterize the response of the nucleon to an external electromagnetic field. Accurate determination of the nucleon polarizabilities is an ongoing effort that has lately received considerable attention. The polarizabilities may be accessed from the differential cross section for Compton scattering on the nucleon. Several exhaustive theoretical reviews that summarize and interpret the available Compton data have been published over the last few decades. An approach to describing Compton scattering on light nuclei based upon dispersion relations is discussed in Refs.~\cite{Drechsel200399, Schumacher2005567}, whereas an effective field-theory approach is presented in Ref.~\cite{griesshammer2012}.

The proton polarizabilities are extracted from the cross section for Compton scattering on hydrogen, while for neutron polarizabilities, a light nuclear target is required since no feasible free-neutron target exists. Unfolding the neutron polarizabilities from the coherent Compton-scattering signal on a light nucleus is more complicated because of the interactions due to the additional nucleon(s). Furthermore, the world data set for Compton scattering on hydrogen is vastly larger compared to the Compton data set for light nuclear targets, such as deuterium and helium~\cite{griesshammer2012}. Consequently, the proton polarizabilities are known considerably more accurately. The latest determination of the electric and magnetic polarizabilities $\alpha$ and $\beta$ of the proton yielded~\cite{mcgovern2012}
\begin{align}
  &\alpha_p &=& 10.65 \pm 0.35(\text{stat}) \pm 0.2(\text{BSR}) \pm 0.3(\text{th}), \\ 
  &\beta_p  &=& \phantom{0}3.15  \mp 0.35(\text{stat}) \pm 0.2(\text{BSR}) \mp 0.3(\text{th}).
\end{align}
The uncertainty due to the Baldin sum rule (BSR)~\cite{baldin1960}, which is based on the optical theorem, is indicated separately from the statistical and theoretical uncertainties. Both $\alpha$ and $\beta$ are expressed in the conventional unit $10^{-4}$~${\rm fm^3}$, which is used throughout this article.

The first modern measurements of the neutron polarizabilities via Compton scattering on the deuteron below pion-production threshold were performed in Illinois~\cite{lucas1994} and Saskatchewan~\cite{hornidge2000}. These were shortly followed up by an experiment in Lund~\cite{lundin2003}. In parallel, extractions of the neutron polarizabilities via quasifree Compton scattering on the deuteron were performed as described in Refs.~\cite{PhysRevLett.85.1388,PhysRevLett.88.162301}. Both experiments used incident photon energies $E_\gamma>200$~MeV and data was interpreted using dispersion theory. The most accurate determination of the neutron polarizabilities to date was performed in Refs.~\cite{PhysRevLett.113.262506,PhysRevC.92.025203,Griesshammer:2015ahu} through combining new Compton scattering on the deuteron data with the world data set, leading to
\begin{align}
  &\alpha_n &=& 11.55 \pm 1.25(\text{stat}) \pm 0.2(\text{BSR}) \pm 0.8(\text{th}), \\ 
  &\beta_n  &=& \phantom{0}3.65  \mp 1.25(\text{stat}) \pm 0.2(\text{BSR}) \mp 0.8(\text{th}).
\end{align}
Comparison of the proton and the neutron polarizabilities indicates that knowledge of the latter is substantially poorer, with larger statistical and theoretical uncertainties. The sensitivity of the cross section for Compton scattering to the polarizabilities increases with the incident photon energy~\cite{griesshammer2012}, which makes measurements above the pion-production threshold a possible means to improve the accuracy of the extracted neutron polarizabilities. However, this introduces new experimental and theoretical challenges. Experimentally, the separation of coherent Compton scattering on the deuteron from scattering off the bound nucleons becomes more difficult, as it relies on photon detectors with an energy resolution of the order of the deuteron break-up energy. Additionally, the strong background from radiative capture of photoproduced $\pi^-$ in deuterium overwhelms the much weaker signal from the Compton-scattering reaction. This precludes measurements at energies and angles where the Compton-scattering signal and the $\pi^-$-capture signal overlap kinematically. Theoretically, the interpretation of the Compton cross section becomes more complicated due to the opening of the pion-production channels and the growing influence of the $\Delta$-resonance. These challenges need to be addressed for an improved determination of the neutron polarizabilities from Compton data at these energies. In this article, the first measurement of the cross section for Compton scattering on the deuteron above pion-production threshold is presented. The data presented below are anticipated to stimulate further development of the theory and offer constraints for future calculations and experiments.

\begin{figure*}[ht]
  \begin{minipage}{.45\textwidth}

    \centering
    \scalebox{0.55}{

\begin{tikzpicture}

\pgfmathsetmacro{\SCALE}{200}  
\pgfmathsetmacro{\TargetDiameter}{68/\SCALE}  
\pgfmathsetmacro{\TargetZ}{150/\SCALE}    

\filldraw[fill=blue] (-\TargetZ/2, -\TargetDiameter/2) rectangle (\TargetZ/2, \TargetDiameter/2);

\pgfmathsetmacro{\BuniDiameter}{761.7/\SCALE}  
\pgfmathsetmacro{\BuniZ}{559/\SCALE}    
\pgfmathsetmacro{\BuniCollDiameter}{150/\SCALE}    
\pgfmathsetmacro{\BuniCollimatorZ}{96/\SCALE}    
\pgfmathsetmacro{\BuniCollToCryst}{318/\SCALE}    
\pgfmathsetmacro{\BuniTargetToColl}{381/\SCALE}

\pgfmathsetmacro{\CatsDiameter}{703/\SCALE}  
\pgfmathsetmacro{\CatsZ}{635/\SCALE}    
\pgfmathsetmacro{\CatsCollDiameter}{138/\SCALE}    
\pgfmathsetmacro{\CatsCollimatorZ}{210/\SCALE}    
\pgfmathsetmacro{\CatsCollToCryst}{450/\SCALE}    
\pgfmathsetmacro{\CatsTargetToColl}{297/\SCALE}    
\pgfmathsetmacro{\CatsInnerCollZ}{146/\SCALE}    
\pgfmathsetmacro{\CatsInnerCollOD}{260/\SCALE}    
\pgfmathsetmacro{\CatsInnerCollIDf}{138/\SCALE}    
\pgfmathsetmacro{\CatsInnerCollIDb}{187/\SCALE}    

\begin{scope}[rotate = -60, xshift = 5.32cm]

  \filldraw[fill=green!50!blue] (-\CatsZ/2, -\CatsDiameter/2) rectangle (\CatsZ/2, \CatsDiameter/2); 
  \node at (0, 0) {\normalsize CATS};
  \filldraw[fill=black] (-\CatsZ/2-\CatsCollToCryst, \CatsCollDiameter/2) rectangle (-\CatsZ/2-\CatsCollToCryst+\CatsCollimatorZ, \CatsDiameter/2);
  \filldraw[fill=black] (-\CatsZ/2-\CatsCollToCryst, -\CatsCollDiameter/2) rectangle (-\CatsZ/2-\CatsCollToCryst+\CatsCollimatorZ, -\CatsDiameter/2);

  \draw[fill=black] (-\CatsZ/2-\CatsInnerCollZ, -\CatsInnerCollOD/2) -- (-\CatsZ/2-\CatsInnerCollZ, -\CatsInnerCollIDf/2) -- (-\CatsZ/2, -\CatsInnerCollIDb/2) -- (-\CatsZ/2, -\CatsInnerCollOD/2) -- cycle;

  \draw[fill=black] (-\CatsZ/2-\CatsInnerCollZ, \CatsInnerCollOD/2) -- (-\CatsZ/2-\CatsInnerCollZ, \CatsInnerCollIDf/2) -- (-\CatsZ/2, \CatsInnerCollIDb/2) -- (-\CatsZ/2, \CatsInnerCollOD/2) -- cycle;

  \draw[<->] (-\CatsZ/2-\CatsCollToCryst+\CatsCollimatorZ, -\CatsDiameter/2) -- (-\CatsZ/2, -\CatsDiameter/2);
  \node[rotate=-60] at (-2.2, -2.0) {\normalsize 240 mm};

  \draw[<->] (-\CatsZ/2-\CatsCollToCryst, \CatsDiameter/2+0.2) -- (-\CatsZ/2-\CatsCollToCryst+\CatsCollimatorZ, \CatsDiameter/2+0.2);
  \node[rotate=-60] at (-3.3, 2.2) {\normalsize 210 mm};

  \draw[<->] (\CatsZ/2 + 0.2, -\CatsDiameter/2) -- (\CatsZ/2 + 0.2, \CatsDiameter/2);
  \node[rotate=30] at (\CatsZ-1, 0) {\normalsize 703 mm};

  \draw[<->] (-\CatsZ/2, \CatsDiameter/2 + 0.2) -- (\CatsZ/2, \CatsDiameter/2 + 0.2);
  \node[rotate=-60] at (0, 2.3) {\normalsize 635 mm};

  \draw[<->] (-\CatsZ/2-\CatsCollToCryst+\CatsCollimatorZ, -\CatsCollDiameter/2) -- (-\CatsZ/2-\CatsCollToCryst+\CatsCollimatorZ, \CatsCollDiameter/2);
  \node[rotate=30] at (-\CatsCollToCryst-0.3, 0) {\normalsize 138 mm};

  \draw[<->] (-\CatsZ/2-\CatsInnerCollZ, -\CatsInnerCollIDf/2) -- (-\CatsZ/2-\CatsInnerCollZ, \CatsInnerCollIDf/2);

  \draw (-\CatsZ/2, -\CatsInnerCollIDb/2) -- (-\CatsZ/2+0.2, -\CatsInnerCollIDb/2);
  \draw (-\CatsZ/2, \CatsInnerCollIDb/2) -- (-\CatsZ/2+0.2, \CatsInnerCollIDb/2);
  \draw[<->] (-\CatsZ/2+0.2, -\CatsInnerCollIDb/2) -- (-\CatsZ/2+0.2, \CatsInnerCollIDb/2);
  \node[rotate=30] at (-\CatsCollToCryst+1.1, 0) {\normalsize 187 mm};

  \draw[dotted] (-\CatsZ/2, -\CatsInnerCollOD/2) -- (-\CatsZ/2+0.8, -\CatsInnerCollOD/2);
  \draw[dotted] (-\CatsZ/2, \CatsInnerCollOD/2) -- (-\CatsZ/2+0.8, \CatsInnerCollOD/2);
  \draw[<->] (-\CatsZ/2+0.8, -\CatsInnerCollOD/2) -- (-\CatsZ/2+0.8, \CatsInnerCollOD/2);
  \node[rotate=30] at (-\CatsCollToCryst+1.7, 0) {\normalsize 260 mm};

  \draw[<->] (-\CatsZ/2-\CatsInnerCollZ, \CatsInnerCollOD/2+0.2) -- (-\CatsZ/2, \CatsInnerCollOD/2+0.2);
  \node[rotate=30] at (-\CatsCollToCryst+0.3, 1.5) {\normalsize 146 mm};

\end{scope}

\draw[rotate=-60,<->] (\CatsTargetToColl, 0) -- (0, 0);
\node[rotate=-60,text width=2cm, align=center] at (-0.4,-1.1) {\normalsize 297 mm \\ (335 mm)};

\draw[loosely dotted, ->] (0, 0) -- (4, 0)            
node[above=0.3cm, pos=0.8] {\normalsize z-axis};

\draw[rotate=-60, loosely dotted] (0, 0) -- (\CatsTargetToColl+\CatsCollToCryst, 0);  

\draw (1.2, 0) arc (0:-60:1.2cm);                      
\node at (0.87, -0.5) {\normalsize 60\textdegree};

\draw[decorate, decoration={snake, segment length=3mm, amplitude=1mm}, -stealth] (-4, 0) -- (-\TargetZ/2,0) 
node[above=0.2cm, pos=0.4] {\normalsize $\gamma$};

\filldraw[gray] (-4.5, 0.2) rectangle (-4.4, 0.7);
\filldraw[gray] (-4.5, -0.2) rectangle (-4.4, -0.7);
\draw[<-] (-4.4, -0.45) -- (-3.4, -0.45);
\node[text width=2cm, align=left] at (-3.3, -1.1)  {\normalsize 3720 mm from $\rm {}^2H$ target};
\draw[<->] (-4.45, 0.2) -- (-4.45, -0.2);
\node at (-5.2, 0) {\normalsize 19 mm};

\end{tikzpicture}

     }
    \caption{(Color online) A drawing of the experimental setup. Gray - beam collimator; wavy line - photon beam; blue box - deuterium target; black regions - CATS front and inner collimator; green box - CATS scintillators (NaI(Tl) and plastics).}
    \label{fig:LundFloorPlan}

  \end{minipage}
  \hfill  
  \begin{minipage}{.45\textwidth}

    \centering
    \scalebox{0.8}{
      
\begin{tikzpicture}

\pgfmathsetmacro{\SCALE}{200}
\pgfmathsetmacro{\CatsCoreR}{133.5/\SCALE}  
\pgfmathsetmacro{\CatsSegR}{\CatsCoreR + 108/\SCALE}  
\pgfmathsetmacro{\CatsPlasR}{\CatsSegR + 100/\SCALE}  

\coordinate (c1) at (0,0);

\draw[pattern=north east lines] (0,0) circle (\CatsCoreR);

\draw[pattern=north east lines, pattern color=green!60]
($(c1) + (60:\CatsCoreR)$) arc (60:120:\CatsCoreR) --
($(c1) + (120:\CatsSegR)$) arc (120:60:\CatsSegR) -- cycle;

\draw[pattern=north east lines]
($(c1) + (60+1*60:\CatsCoreR)$) arc (60+1*60:120+1*60:\CatsCoreR) --
($(c1) + (120+1*60:\CatsSegR)$) arc (120+1*60:60+1*60:\CatsSegR) -- cycle;

\draw[pattern=north east lines]
($(c1) + (60+2*60:\CatsCoreR)$) arc (60+2*60:120+2*60:\CatsCoreR) --
($(c1) + (120+2*60:\CatsSegR)$) arc (120+2*60:60+2*60:\CatsSegR) -- cycle;

\draw[pattern=north east lines, pattern color=green!60]
($(c1) + (60+3*60:\CatsCoreR)$) arc (60+3*60:120+3*60:\CatsCoreR) --
($(c1) + (120+3*60:\CatsSegR)$) arc (120+3*60:60+3*60:\CatsSegR) -- cycle;

\draw[pattern=north east lines]
($(c1) + (60+4*60:\CatsCoreR)$) arc (60+4*60:120+4*60:\CatsCoreR) --
($(c1) + (120+4*60:\CatsSegR)$) arc (120+4*60:60+4*60:\CatsSegR) -- cycle;

\draw[pattern=north east lines]
($(c1) + (60+5*60:\CatsCoreR)$) arc (60+5*60:120+5*60:\CatsCoreR) --
($(c1) + (120+5*60:\CatsSegR)$) arc (120+5*60:60+5*60:\CatsSegR) -- cycle;

\draw[fill=green!60]
($(c1) + (-90:\CatsSegR)$) arc (-90:-90+1*72:\CatsSegR) --
($(c1) + (-90+1*72:\CatsPlasR)$) arc (-90+1*72:-90:\CatsPlasR) -- cycle;

\draw[fill=white]
($(c1) + (-90+1*72:\CatsSegR)$) arc (-90+1*72:-90+2*72:\CatsSegR) --
($(c1) + (-90+2*72:\CatsPlasR)$) arc (-90+2*72:-90+1*72:\CatsPlasR) -- cycle;

\draw[fill=green!60]
($(c1) + (-90+2*72:\CatsSegR)$) arc (-90+2*72:-90+3*72:\CatsSegR) --
($(c1) + (-90+3*72:\CatsPlasR)$) arc (-90+3*72:-90+2*72:\CatsPlasR) -- cycle;

\draw[fill=white]
($(c1) + (-90+3*72:\CatsSegR)$) arc (-90+3*72:-90+4*72:\CatsSegR) --
($(c1) + (-90+4*72:\CatsPlasR)$) arc (-90+4*72:-90+3*72:\CatsPlasR) -- cycle;

\draw[fill=white]
($(c1) + (-90+4*72:\CatsSegR)$) arc (-90+4*72:-90+5*72:\CatsSegR) --
($(c1) + (-90+5*72:\CatsPlasR)$) arc (-90+5*72:-90+4*72:\CatsPlasR) -- cycle;

\draw[dotted] (-\CatsCoreR, 0) -- (-\CatsCoreR, -\CatsPlasR - 0.5);
\draw[dotted] (\CatsCoreR, 0) -- (\CatsCoreR, -\CatsPlasR - 0.5);
\draw[<->] (-\CatsCoreR, -\CatsPlasR - 0.5) -- (\CatsCoreR, -\CatsPlasR - 0.5) node[below,pos=0.5] {\normalsize 267~mm};

\draw[dotted] (-\CatsSegR, 0) -- (-\CatsSegR, -\CatsPlasR - 1);
\draw[dotted] (\CatsSegR, 0) -- (\CatsSegR, -\CatsPlasR - 1);
\draw[<->] (-\CatsSegR, -\CatsPlasR - 1) -- (\CatsSegR, -\CatsPlasR - 1) node[below,pos=0.5] {\normalsize 483~mm};

\draw[->] (-0.5, 2) -- (0.5, -2);

\end{tikzpicture}

     }
    \vspace{1.5mm}
    \caption{(Color online) Cross-sectional view of the CATS detector. Cosmic-ray muon (downgoing arrow) events that caused a signal in opposing annulus segments (green) were selected for monitoring PMT gain instabilities. NaI(Tl) crystals are striped to distinguish them from plastic scintillators.}
    \label{fig:NaI_CSview}

  \end{minipage}
\end{figure*}

The experiment was performed at the former Tagged-Photon Facility~\cite{adler2012} located at the MAX IV Laboratory~\cite{eriksson2014} in Lund, Sweden. A pulse-stretched electron beam~\cite{lindgren2002} with an energy of ${\sim}190$~MeV was used to produce quasi-monoenergetic photons via the bremsstrahlung-tagging technique~\cite{adler1990,adler1997}. The photons were tagged in the energy range 140 -- 160~MeV and the energies of the photons were determined with a resolution of ${\sim}0.6$~MeV by the 64-channel SAL focal-plane (FP) hodoscope~\cite{vogt1993}. The arrival times of post-bremsstrahlung recoil electrons at the hodoscope were digitized with multi-hit time-to-digital (TDC) converters. Time-normalized scalers provided the post-bremsstrahlung recoil electron counting rate, which was typically $0.1-1$~MHz per FP channel. Daily tagging-efficiency measurements were employed to account for the tagged photons that were lost in the beam-collimation process. To that end, a ${\sim}100$\% efficient Pb-glass detector was raised into the beam to count the bremsstrahlung photons that passed through the collimator into the experimental hall. The tagging efficiency $\epsilon$ for each FP channel was determined as
\begin{equation}
  \epsilon_{\rm ch} = N_{\rm Pb}^{\rm ch}/N_{\rm FP}^{\rm ch},
\end{equation}
where $N_{\rm Pb}^{\rm ch}$ was the number of bremsstrahlung photons that entered the experimental hall (detected by the Pb-glass detector) and $N_{\rm FP}^{\rm ch}$ was the number of post-bremsstrahlung recoil electrons incident on the FP channel. The tagging-efficiency measurements were performed with a low-intensity photon beam, which enabled unambiguous correlation of a photon detected in the Pb-glass with an electron detected in a specific FP channel. The tagging efficiencies varied slightly along the entire FP detector, but had a mean value of ${\sim}23$\% with a systematic uncertainty of ${\sim}2$\%.

The tagged-photon beam was incident on a cylindrical Kapton vessel that contained liquid deuterium. The vessel was 170~mm long aligned along the beam axis, with a diameter of 68~mm and 120~$\si{\micro\metre}$ thick walls. The density of the liquid deuterium was monitored throughout the experiment and was stable at ${\rho_D = (0.163 \pm 0.001)}$~g/$\rm cm^3$. A large NaI(Tl) spectrometer named CATS~\cite{Hunger1997385} was set up at a laboratory angle of $\theta=60^\circ$ to detect Compton-scattered photons. Two other NaI(Tl) spectrometers recorded events at $\theta=120^\circ$ and $150^\circ$, but in these detectors the Compton signal was largely obscured by the background signal from $\pi^-$ capture. This prevented the extraction of the cross section for Compton scattering at backward angles. However, data from all three detectors were used for the measurement of the cross section for $\pi^-$ photoproduction on the deuteron, which will be discussed in a separate article. 

The CATS spectrometer consisted of a large cylindrical core NaI(Tl) crystal that was surrounded by six optically isolated NaI(Tl) segments. The NaI(Tl) segments were in turn surrounded by an annulus of plastic scintillators that were used for the identification of cosmic-ray muons, which constituted a significant background. The energy resolution of the detector was approximately 3~MeV FWHM for ${\sim}150$~MeV incident photons, which enabled the separation of the coherent deuteron Compton-scattering signal from non-elastic reaction channels. The scintillation was read out by photomultiplier tubes (PMTs) attached to the rear faces of the scintillators. The analog signals from the PMTs were converted to digital format by charge-integrating analog-to-digital converters (QDCs).

Data were recorded on an event-by-event basis. The data acquisition and data analysis software was based on ROOT~\cite{Brun199781} and RooFit~\cite{Verkerke:2003ir} software frameworks. The data acquisition was triggered by an event in the CATS detector, which initiated the read-out of the QDCs and started the multi-hit TDCs. The multi-hit TDC stop signals came from the post-bremsstrahlung recoil electrons striking the FP channels. The QDC signals allowed reconstruction of the detected photon energies, while the FP TDC signals established the coincidence between the post-bremsstrahlung recoil electrons and the scattered photons detected with the CATS spectrometer. The data were collected over three 4-week run periods in June and September 2011 and April 2015. The positioning of the CATS detector relative to the Kapton target and the tagged-photon beam is depicted in Fig.~\ref{fig:LundFloorPlan}. The distance between the target center and the CATS collimator was 297~mm in 2015 and 335~mm in 2011.

The NaI(Tl) spectrometer was calibrated by placing it directly into a low-intensity photon beam and establishing the response as a function of the tagged-photon energy. The overall calibration of the NaI(Tl) detector and the tagged-photon energies was confirmed with an accuracy of $\pm0.4$~MeV by reconstructing the ${\sim}130~$MeV photon energy resulting from the radiative capture of $\pi^-$ in deuterium~\cite{gabioud1979} and the location of the $\pi^-$ photoproduction threshold on the FP detector. The gain instabilities of the PMTs over the run periods were corrected for by monitoring the QDC peak locations resulting from cosmic-ray muons traversing the detector on a run-by-run basis. The annulus NaI(Tl) and plastic scintillators were used to select cosmic-ray muons with similar track lengths through the core crystal, which served to suppress the variation of the PMT signals due to varying muon paths. The selection of these events is illustrated in Fig.~\ref{fig:NaI_CSview}.

The differential cross section for Compton scattering was computed according to
\begin{equation}
  \frac{d\sigma}{d\Omega} = \frac{Y}{\Omega_{\rm eff} \cdot N_\gamma \cdot \kappa_{\rm eff}}. \label{eq:ComptonCrossSec}
\end{equation}
In Eq.~\eqref{eq:ComptonCrossSec}, $Y$ is the yield of Compton-scattered photons, $\Omega_{\rm eff}$ is the detector acceptance, $N_\gamma$ is the number of tagged photons incident on the target and $\kappa_{\rm eff}$ is the effective target thickness. For the extraction of the yield from the experimental signal, a missing energy was defined as
\begin{equation}
E_{\rm miss} = E_{\rm det} - E_\gamma'. \label{eq:Emiss}
\end{equation}
In Eq.~\eqref{eq:Emiss}, $E_{\rm det}$ is the energy detected by CATS and $E_\gamma'$ is the expected energy of the Compton-scattered photon, calculated via two-body kinematics from the tagged-photon energy $E_\gamma$, the deuteron mass $M_D$ and the scattering angle $\theta$ according to
\begin{equation}
  E_\gamma' = \frac{M_D E_\gamma}{M_D + E_\gamma(1 - \cos\theta)}. \label{eq:ComptonScattedGammaE}
\end{equation}

\begin{figure*}[t]
  \begin{minipage}{.45\textwidth}
    \centering
    \includegraphics[scale=0.4]{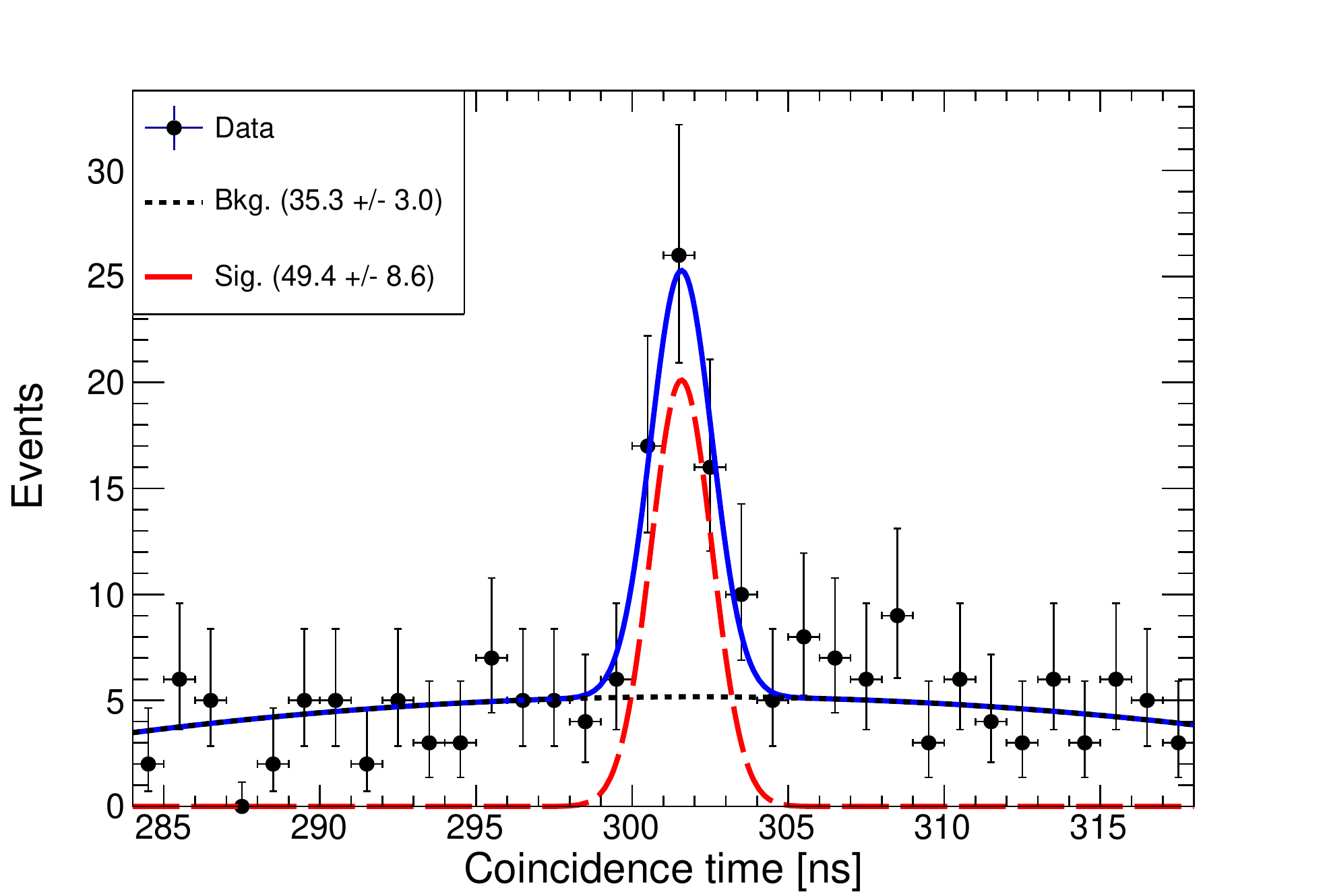}
    \caption{(Color online) Extraction of the Compton yield from a fit to the coincidences within the cut ${E_{\rm miss} \in [-2,\:3]}$~MeV. Filled black circles - data points; long-dashed red - model for true coincidences; short-dashed black - model for background from random coincidences; solid blue - the fit result. The error bars indicate statistical uncertainties.}
    \label{fig:timing_fit}
  \end{minipage}
  \hfill  
  \begin{minipage}{.45\textwidth}
    \centering
    \includegraphics[scale=0.4]{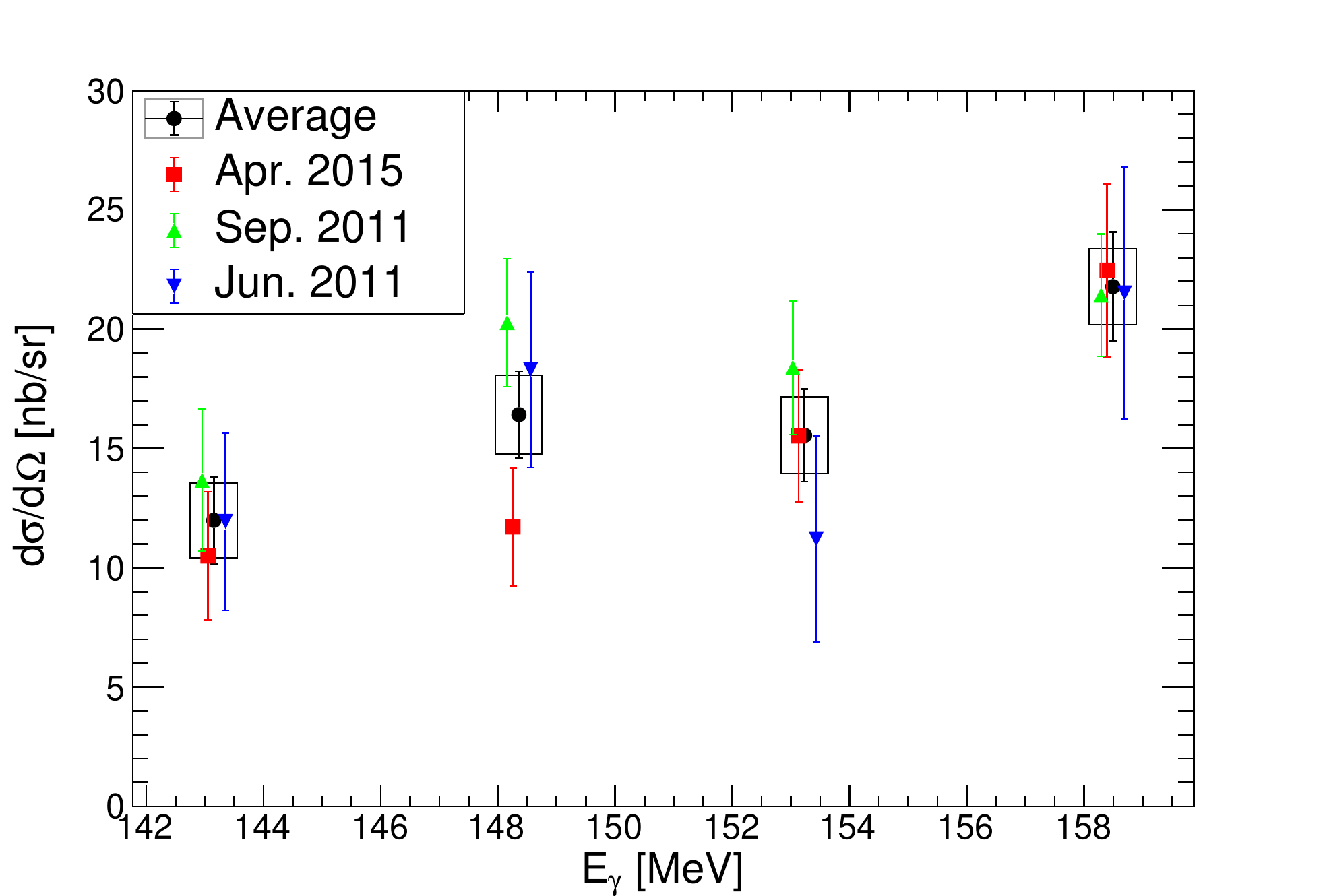}
    \caption{(Color online) Measured cross section for Compton scattering on the deuteron. The measurements from three different run periods are indicated with statistical uncertainties. The average of the three measurements (black circles) are shown with both statistical (error bars) and systematic (error boxes) uncertainties.}
    \label{fig:cats_dcs}
  \end{minipage}
\end{figure*}

\setlength\extrarowheight{2pt}

\begin{table*}[ht]
\centering
\textbf{Measured Compton cross section on the deuteron}\par\medskip
\begin{tabular}{ | c || c | c | c || c | }
  \hline
  $E_\gamma$~[MeV] & Apr.~15~$d\sigma/d\Omega$~[nb/sr] & Sep.~11~$d\sigma/d\Omega$~[nb/sr] & Jun.~11~$d\sigma/d\Omega$~[nb/sr] & Ave.~$d\sigma/d\Omega$~[nb/sr] \\ \hline
  143.2 & $10.5 \pm 2.7 \pm 0.9$ & $13.7 \pm 3.0 \pm 2.1$ & $11.9 \pm 3.7 \pm 1.8$ & $12.0 \pm 1.8 \pm 1.6$ \\ \hline
  148.4 & $11.7 \pm 2.5 \pm 0.6$ & $20.3 \pm 2.7 \pm 2.5$ & $18.3 \pm 4.1 \pm 2.1$ & $16.4 \pm 1.8 \pm 1.7$ \\ \hline
  153.2 & $15.5 \pm 2.8 \pm 0.8$ & $18.4 \pm 2.8 \pm 2.3$ & $11.2 \pm 4.3 \pm 1.7$ & $15.6 \pm 1.9 \pm 1.6$ \\ \hline
  158.5 & $22.5 \pm 3.6 \pm 1.4$ & $21.4 \pm 2.5 \pm 1.7$ & $21.5 \pm 5.3 \pm 1.8$ & $21.8 \pm 2.3 \pm 1.6$ \\ \hline
\end{tabular}
\caption{Measured differential cross section for Compton scattering on the deuteron at $\theta=60^\circ$. The first uncertainty is statistical and the second uncertainty is systematic. A statistically weighted average of the three measurements is reported in the right column.}
\label{tab:cats_dcs}
\end{table*}

\noindent The Compton yield was determined by events that were inside the cut $E_{\rm miss} \in [-2,\:3]$~MeV and were in coincidence with the post-bremsstrahlung electron used to tag the photon. The background from scattering off bound nucleons $\gamma+{\rm {}^2H} \rightarrow \gamma' +{\rm n + p}$ was strongly suppressed by the cut on $E_{\rm miss}$ at $-2$~MeV. GEANT4~\cite{Geant4} simulations confirmed the inelastic contribution to the Compton signal to be~$\lesssim2$\%. The contamination from scattering from the Kapton vessel was also investigated with a GEANT4 simulation and was determined to be $(1.5 \pm 1)$\%. The combined uncertainty from the Kapton and the inelastic contamination was estimated to be 3\% by adding the two contributions in quadrature.

The FP channels were grouped in four energy bins of approximately 5~MeV width. For each energy bin, a coincidence histogram with events that satisfied the cut on $E_{\rm miss}$ was filled. As the $\pi^-$ background was eliminated and the inelastic and Kapton scattering events were strongly suppressed by the $E_{\rm miss}$ cut, the remaining coincidences originated predominantly from the Compton scattering events. This enabled the extraction of the yields directly from fits to the coincidence spectra, as illustrated in Fig.~\ref{fig:timing_fit}. The background from random coincidences had a non-trivial time structure due to a time modulation of the electron-beam intensity, related to the pulse-stretching and beam-extraction apparatus~\cite{jebali2013}. Two background models were used for the yield-extraction fits. In one case, the background was represented by a polynomial and in the other case the background was estimated with the Sensitive Nonlinear Iterative Peak (SNIP) clipping algorithm~\cite{MORHAC1997113,HAMPTON1994280}. With each background model, the fits were performed several times, employing a different fit range around the coincidence peak and using 0.5~ns and 1~ns binning for the data. The different settings and background models led to a total of 32 fits per coincidence spectrum. The standard deviation of the resulting random-subtracted Compton yields was used as an estimate for the systematic uncertainty. The time modulation of the electron-beam intensity was smaller in 2015 and the signal-to-background ratio was less favorable in the 2011 data. As a result, the yields in the first two run periods were more sensitive to the variations in the fit settings. The systematic uncertainties of the yields varied from 6\%~--~15\% in June/September 2011 and from 3\%~--~8\% in April 2015, depending on the energy bin. However, the statistical uncertainties of the yields dominated the overall uncertainties of the cross sections, which ranged from 12\%~--~38\% for different run periods.

The detector acceptance $\Omega_{\rm eff}$ was determined from a GEANT4 simulation. First, the detector response for the simulation was determined by matching the Monte Carlo in-beam data to the experimental in-beam data. To that end, the simulated data were convoluted with a Gaussian function and fitted to the experimental data. This determined the parameters of the convoluting Gaussian that encoded the detector response and could subsequently be used to smear the detected energy of simulated events on an event-by-event basis. A similar approach was used in Ref.~\cite{PhysRevC.92.025203}. Secondly, Compton scattering from the liquid-deuterium target was simulated, where the scattering angle was sampled from a phase-space distribution. The ratio of Monte Carlo events that entered the CATS detector and satisfied the $E_{\rm miss}$ cut to the total number of simulated events determined the acceptance. The different positioning of the detector in different run periods led to moderately different acceptances, with typical values ${\sim}26$~msr in 2015 and ${\sim}23$~msr in 2011. The systematic uncertainty of the acceptance was determined to be ${\sim}3$\% by varying the detector and the target positioning in the simulation by the positioning uncertainty in the experimental hall and calculating the change in the acceptance.

The tagged-photon flux $N_\gamma$ was established by multiplying the counts in the FP hodoscope channels by the measured tagging efficiencies. The systematic uncertainty in the tagged-photon flux was dominated by the ${\sim}2$\% uncertainty in the tagging efficiency.

The average flight path of the photons through the target was combined with the target density to calculate the effective target thickness
\begin{equation}
  \kappa_{\rm eff} = (8.14 \pm 0.10) \cdot 10^{23}\:\rm{nuclei/cm^2}.
\end{equation}
The ${\sim}1.2$\% uncertainty originates predominantly from the variance of the photon flight path through the liquid deuterium, caused by the geometry of the beamline and the Kapton container. The average photon flight path and its variance were determined from a dedicated GEANT4 simulation of the geometry of the beamline and the target. The same target was used in the experiment of Ref.~\cite{PhysRevC.92.025203} and the effective target thicknesses are in excellent agreement.

\begin{table}[t]
\begin{tabular}{ |c c C{1.5cm} C{1.5cm} R{1.5cm}| }
  \hline
  quantity          & source      & un-correlated  & $E_\gamma$ correlated & $E_\gamma$ and run period correlated \\ \hline\hline
  $Y$               & fit         &               &             & 3\% -- 15\%   \\
                    & inelastic   &               &             & $\lesssim2$\% \\
                    & kapton      &               &             & 1\%           \\
  $\rm \Omega_{eff}$ & positioning &               & ${\sim}3$\% &               \\
  $N_\gamma$         & tagg.~eff.  & ${\sim}2$\%   &             &               \\
  $\kappa_{\rm eff}$ & geometry    &               &             & 1.2\%         \\ \hline\hline
\end{tabular}
\caption{Summary of systematic uncertainties. The dominant systematic uncertainties were correlated between energy bins and run periods.}
\label{tab:sys_errs}
\end{table}

The three run periods provided three cross-section measurements that are reported in Table~\ref{tab:cats_dcs}. The measurements from different run periods are mostly consistent within statistical uncertainties and the extracted cross section increases with incident photon energy, which is the expected trend above the pion-production threshold~\cite{griesshammer2012}. The statistically weighted average of the three results is reported in the right column of Table~\ref{tab:cats_dcs}. The leading sources of systematic uncertainties are summarized in Table~\ref{tab:sys_errs}. The systematic uncertainties were dominated by the uncertainties in the yield extraction, which were correlated between run periods. Therefore, in the calculation of the systematic uncertainties of the combined results complete correlation was assumed, which led to the most conservative (largest) combined systematic uncertainties. However, the statistical uncertainties are larger than the systematic uncertainties with typical values between 10\% -- 15\% for the combined results. Figure~\ref{fig:cats_dcs} presents the data of Table~\ref{tab:cats_dcs}. The present cross sections are of similar magnitude to the published data below pion-production threshold~\cite{PhysRevC.92.025203} and also quite similar to the world data set for Compton scattering on the proton at similar energies~\cite{griesshammer2012}. A rigorous theoretical calculation is required for a quantitative comparison to previous measurements and for the incorporation of the present results to the data base for the determination of the neutron polarizabilities.

In summary, a first measurement of the cross section for Compton scattering on the deuteron above the pion-production threshold has been performed. These new data are intended to spur the extension of theoretical approaches for deuteron Compton scattering to energies above the pion threshold. However, further measurements with wider angular coverage and higher statistical significance are required to substantially constrain $\alpha_n$ and $\beta_n$.

It has been shown that the differential cross section for Compton scattering on ${\rm {}^3He}$ also has significant sensitivity to the neutron polarizabilities~\cite{PhysRevLett.98.232303,Shukla200998}. This has led to proposals to measure the differential cross section for elastic Compton scattering on ${\rm{}^3He}$ at HI$\vec{\gamma}$S~\cite{HIGSexp} and at Mainz~\cite{ComptonHe3}. Ref.~\cite{HIGSexp} proposes to measure the cross section at incident photon energies of 100~MeV and 120~MeV at fixed angles by detecting the scattered photons with large NaI(Tl) detectors. Below pion threshold the high resolution of the large detectors will suffice to cleanly separate the elastic events from the breakup channels. Ref.~\cite{ComptonHe3} proposes to measure the cross section at incident photon energies from 50~MeV~--~200~MeV with ${\sim}4\pi$ angular coverage by using a Helium Gas Scintillator Active Target~\cite{AlJebali2015} in conjunction with the Crystal Ball detector~\cite{PhysRevC.64.055205} at Mainz. The use of the active target will enable the detection of the recoiling helium nucleus, which will be crucial for exploring the region above the pion-production threshold due to the resolution limit of NaI(Tl) calorimeter at Mainz. The proposed experiments could potentially improve upon the statistical uncertainties of the experiment reported here due to the higher cross section for Compton scattering on ${\rm {}^3He}$ compared to ${\rm {}^2H}$ and the substantially wider angular coverage.

The authors acknowledge the support of the staff of the MAX IV Laboratory. The Lund group acknowledges the support from the Swedish Research Council (contracts 40324901 and 80410001) and The Crafoord Foundation (Ref. 20060749). The Glasgow group acknowledges the support from the Scottish Universities Physics Alliance (SUPA), the SUPA Prize Studentship and STFC grants 57071/1 and 50727/1. Finally, support from the DOE grants \mbox{DE--SC0016581}, \mbox{DE--SC0016583}, \mbox{DE--FG02--06ER41422} and the NFS/IRES award 0553467 is acknowledged by the authors in Washington, D.C. and Massachusetts Dartmouth. The authors would like to thank H.W.~Grie{\ss}hammer for constructive discussions.

\bibliography{strandberg_etal}

\end{document}